# Thermoelastic fracturing and buoyancy-driven convection – Surprising sources of longevity for EGS circulation

Mark McClure

mark@resfrac.com

## Introduction

To maximize value from an EGS system, we need to optimize flow rate, well spacing, and well configuration. These engineering decisions will make the difference between a money loser and a cash cow. I've recently run ResFrac simulations of long-term EGS circulation and observed some surprising and intriguing results. Key takeaways:

1. Thermoelastic fracture opening and propagation can have a significant negative effect on the uniformity of flow. On the other hand, interactions between fracture opening and buoyancy-driven fluid circulation cause downward fracture propagation during long-term circulation that greatly improves the thermal longevity of the system.
2. Passive inflow control design can significantly mitigate the negative effect of thermoelastic fracture opening on flow uniformity, while maintaining the positive effects of thermoelastic fracture opening and propagation on flow rate and thermal longevity.
3. Overall, simulations suggest that an EGS doublet with 8000 ft laterals at 475° F – using inflow control at the production well – could sustain electricity generation rates of 8-10 MWe for more than 30 years. Without inflow control, 6-8 MWe over 30 years is possible; however, there is greater risk of uncontrolled thermal breakthrough.

## Background

In Enhanced Geothermal Systems (EGS), fluid is circulated between injection and production wells though stimulated fractures. Fluid heats up as it flows through the fractures because heat conducts inward from the surrounding rock. Over time, the rock around the fractures cools, and the temperature of the produced fluid decreases. Ultimately, the produced fluid will become too low for economic production.

Thermal longevity is critical for EGS economics. Projects are likely to require payout periods of several years. A multiyear payback period may be acceptable for an asset that generates electricity for decades. However, if the produced temperature drops off too rapidly, then the duration of the revenue-generating period will be too short to justify the up-front investment.

Mathematical solutions demonstrate the importance of distributing flow across a large fracture surface area (Gringarten et al., 1975; Doe and McLaren, 2016). To prevent thermal decline at the production well, heat conduction into the fractures must 'keep up' with the rate of fluid flow through the fractures. Unfortunately, heat conduction through rock is relatively slow. At high rate (>100 kg/s), heat conduction requirements imply a huge fracture surface area – on the order of 1e7-1e8 $ft^2$. With modern approaches, this is achievable – multistage plug and perf fracturing techniques are capable of generating enormous surface areas (Raterman et al., 2017; 2019; Gale et al., 2018). Norbeck and Latimer (2023) demonstrate that the flow distribution from plug and perf completion can be distributed fairly evenly across a lateral (Norbeck and Latimer, 2023).

However, the physics of long-term circulation are complex. As fluid injection cools the rock, thermoelastic stress reductions cause crack opening and propagation. This process can have both a negative and a positive impact on thermal breakthrough.

## Simulation setup

In ResFrac, we usually run 'sector' models of a section of the lateral instead of the full well. In EGS simulations, if we 'turn off' thermoelastic stress reductions from cooling, the simulations would run extremely fast, and we could feasibly run full-lateral simulations. For example, the simulation in Figure 1 simulates 30 years of production in a single stage – with no thermoelastic effects – in five minutes of simulation time. However, as shown below, thermoelastic stresses and fracture propagation have a major effect on the results. Including these processes in the simulation greatly increases the computational cost and complexity of the simulation, and so when including the full physics, we cannot yet perform full-lateral simulations.

We need specialized modeling techniques to appropriately describe a full lateral with a 'sector' model in EGS simulations. The well is meshed to the surface. A flow multiplier is used at the heel of the wells so that the vertical sections of the well have the 'full lateral' flow rates (and so, the model will account for friction and heat conduction in the wellbore at the 'full lateral rate'). It is important to consider the 'full lateral' rate in the wellbore friction calculation because it can have a significant impact on BHP. Heat conduction to/from the wells is also included, but it has only a slight effect. The 'flow multiplier' is scaled based on the total lateral length and also from an estimate of the uniformity of flow along the lateral. The flow multiplier penalty for non-uniformity along the full lateral is updated every timestep based on the uniformity of flow within the sector model.

All of the simulations contain a 200 ft stage, scaled for flow along an 8000 ft lateral. The injection wellhead pressure is a constant 2000 psi, and the production wellhead pressure is a constant 600 psi. The injection well is stimulated with a slickwater plug and perf fracturing treatment with 100 mesh proppant. The production well is open-hole and is offset 600 ft laterally. A perforation cluster is placed every 25 ft, with 2 shots per cluster, yielding a strong limited-entry design and relatively uniform fracture geometries along the stage.

# Long-term circulation in a simulation WITHOUT porothermoelastic stress changes

As an initial baseline simulation, Figure 1 shows a simulation with porothermoelastic stresses 'turned off.' In other words, as a model assumption, pressure and temperature changes in the formation do not induce stress changes.

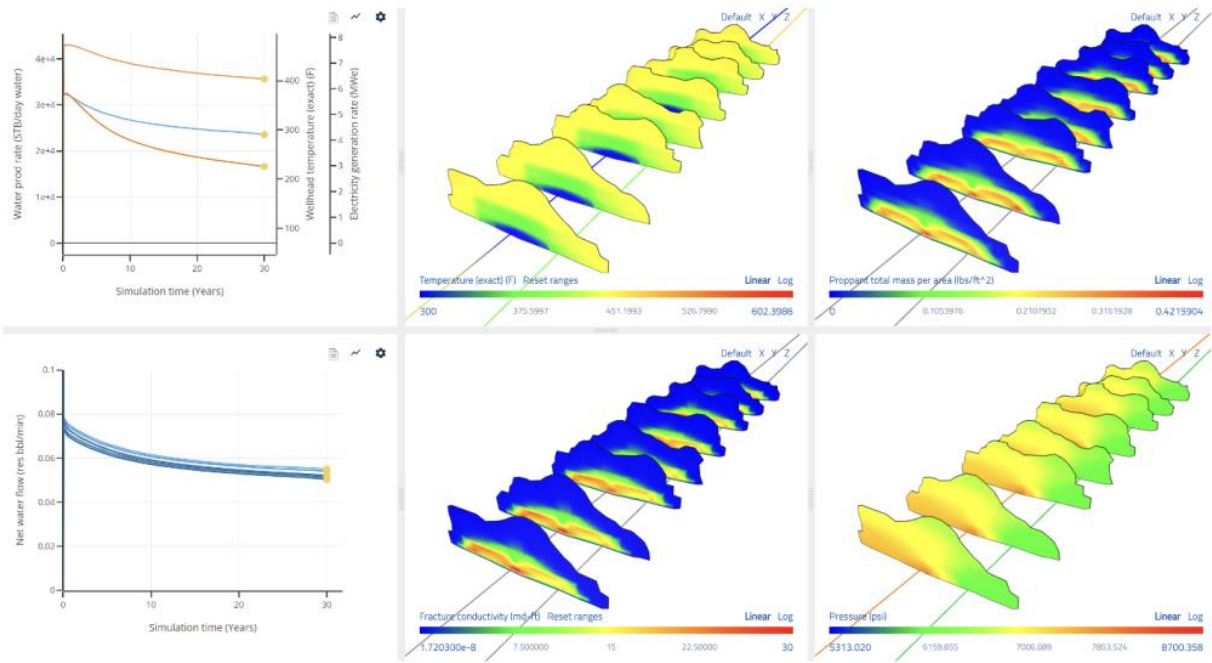

Figure 1: Long-term circulation with porothermoelastic stresses 'turned off'.

As seen in the lower middle panel, the fracture conductivity in the propped regions is around 20 md-ft. In Fervo's recent pilot, they estimated much higher conductivity between the wells – 300-400 md-ft, and so this simulation's conductivity values may be significantly too conservative. The Fervo results are not entirely surprising – in shale, when wells are put on production, fracture conductivity is often over 100 md-ft (Almasoodi et al., 2023), and EGS projects tend to be in lithologies that favor higher fracture conductivity. Conductivity decreases dramatically during long-term pressure depletion, but in an EGS system, long-term injection maintains reservoir pressure, which should prevent a major loss of conductivity. There may be some time-dependent conductivity loss due to gradual crushing, chemical degradation, etc., and this simulation does not consider time-dependent conductivity degradation (although, ResFrac does have the capability to include time-dependent conductivity loss if desired).

To account for the effect of fracture swarming, the 'fractal dimension' of the fractures is set to 0.3 (Section 19.10 from McClure et al., 2023; Acuna, 2020). This has a modest positive impact on the produced temperature.

The initial flow rate is about 32,000 bbl/day (59 kg/s), produced at 475° F, yielding 5.8 MWe of electricity. Over time, the cooling front gradually extends towards the production well, and the produced temperature after 30 years drops to 400° F. The flow distribution is quite uniform in the simulation, probably overoptimistically. Norbeck and Latimer (2023) report a relatively uniform flow distribution, but not as uniform as in this simulation. Rate drops over time because the water in the fracture becomes more viscous as the temperature gradually drops, reaching 23,000 bbl/day (42 kg/s) and 3 MWe of power generation after 30 years. The rate per cluster is initially around 0.08 bbl/min (0.2 kg/s) and drops to around 0.05 bbl/min (0.13 kg/s) by year 30.

# Long-term circulation in a simulation WITH porothermoelastic stress changes

Figure 2 shows results from a simulation with porothermoelastic stress changes 'turned on.' In other words, pressure and temperature changes in the formation *do* induce stress changes. At the conclusion of fracturing, the results are similar to the simulation without porothermoelastic stresses. However, the behavior during long-term circulation is very different.

Over the first few years, the flow rate increases significantly as fracture opening induced by cooling increases the fracture conductivity. To mitigate thermal breakthrough, the maximum permitted production rate is set to 48,000 bbl/day. This rate is reached during year 3. At roughly the same time, the production wellhead temperature experiences a sharp drop and then continues to decline until plateauing around year 9. As the rock cools and stress drops, an 'opening front' develops in each fracture and gradually propagates toward the production well. This is the region of the crack where it has mechanically opened (walls out of contact) because of stress reduction. The abrupt drop in produced wellhead temperature in year 3 occurs when the first fracture opening front reaches the production well, creating a continuous high conductivity pathway between the two wells.

In a real project, the arrival of the first 'opening front' at the production well will be a moment of significant risk. If the short- circuit is too severe, this could cause a severe drop in produced temperature. In this simulation, the temperature drop is abrupt, but not overly severe. There is enough flow from the other fractures to avoid a total short-circuit, and cooling is not sufficiently strong at the 'tip' of the opening front. Nevertheless, if we had used more pessimistic assumptions regarding the uniformity of flow along the lateral, or the initial distribution of fracture conductivity and uniformity, the thermal breakthrough might have been more severe.

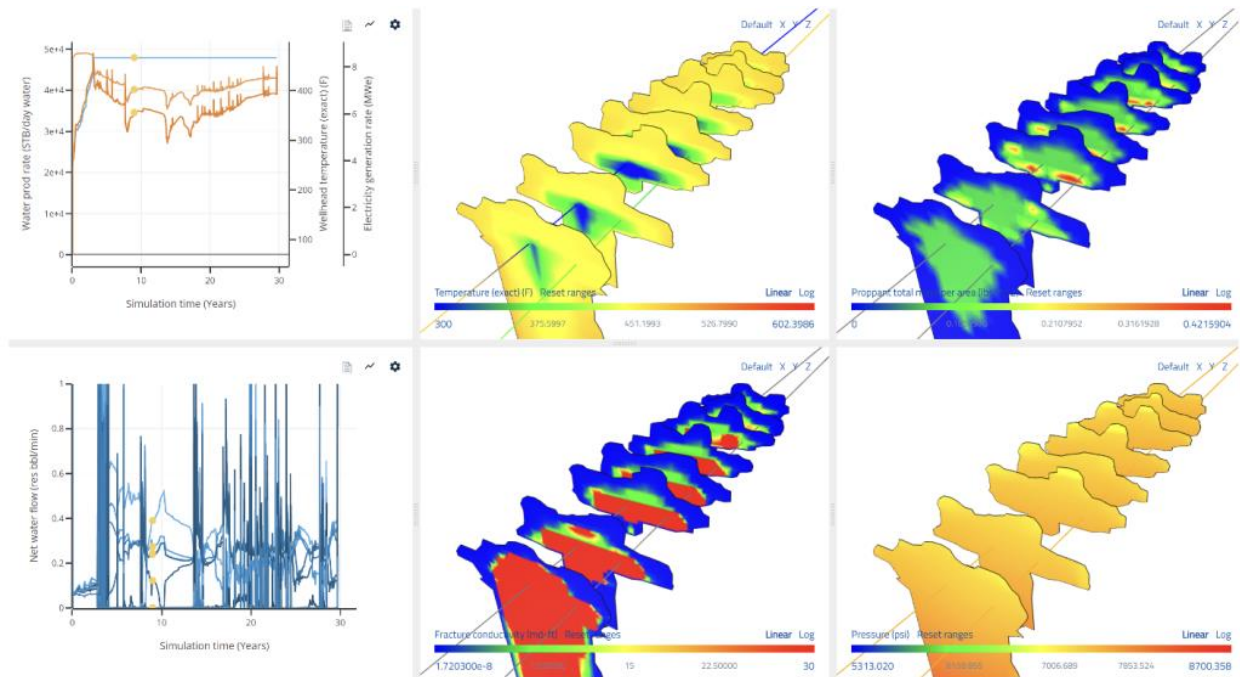

Figure 2: Long-term circulation with porothermoelastic stresses 'turned on'.

The lower left panel in Figure 2 shows flow 'per cluster.' We can see a large increase in flow per cluster during year 3. As noted above, the model adaptively modifies how it scales the sector model to the 'full lateral' based on an estimate of uniformity along the lateral, as evaluated from uniformity within the sector model. When flow becomes more nonuniform within the stage during year 3, the overall 'full lateral' scaling factor drops, forcing even more fluid through the 'sector model' (in order to simulate non-uniformity of flow along the lateral). During this period, the flow rate of four of the fractures drops to near zero; one fracture experiences flow as high as 0.6 bpm, with the other three varying from 0.1 to 0.45 bpm.

During the subsequent years, the decrease in produced temperature surprisingly slows. What is happening? Remarkably, buoyancy-drive convection currents develop inside the 'open' parts of the fractures, greatly improving heat sweep efficiency.

Figure 3 shows the simulation result during year 6. The top middle panel shows temperature; the top left panel shows the porothermoelastic stress change; the top right panel shows water density. In the regions where the fracture is cool, the water density is 15% greater than in the regions with high temperature. The lower left panel shows the fracture conductivity across the model (in log-scale). The regions with mechanically open cracks are red, corresponding to mechanically open, effectively infinite conductivity cracks. In the mechanically open regions, buoyancy-driven convection dominates flow. This can be seen in the middle bottom panel, which shows the vertical volumetric flow rate per lateral area. Fluid flows downward from the injection well, then sweeps back up the sides. This causes a downwardly propagating cooling front, which extends the crack gradually downward over years of circulation. In this image, at year 6, the bottom of the crack is 1500 ft below the injection well; at the start of circulation, the bottom of the crack had been only a few hundred ft below the injection well.

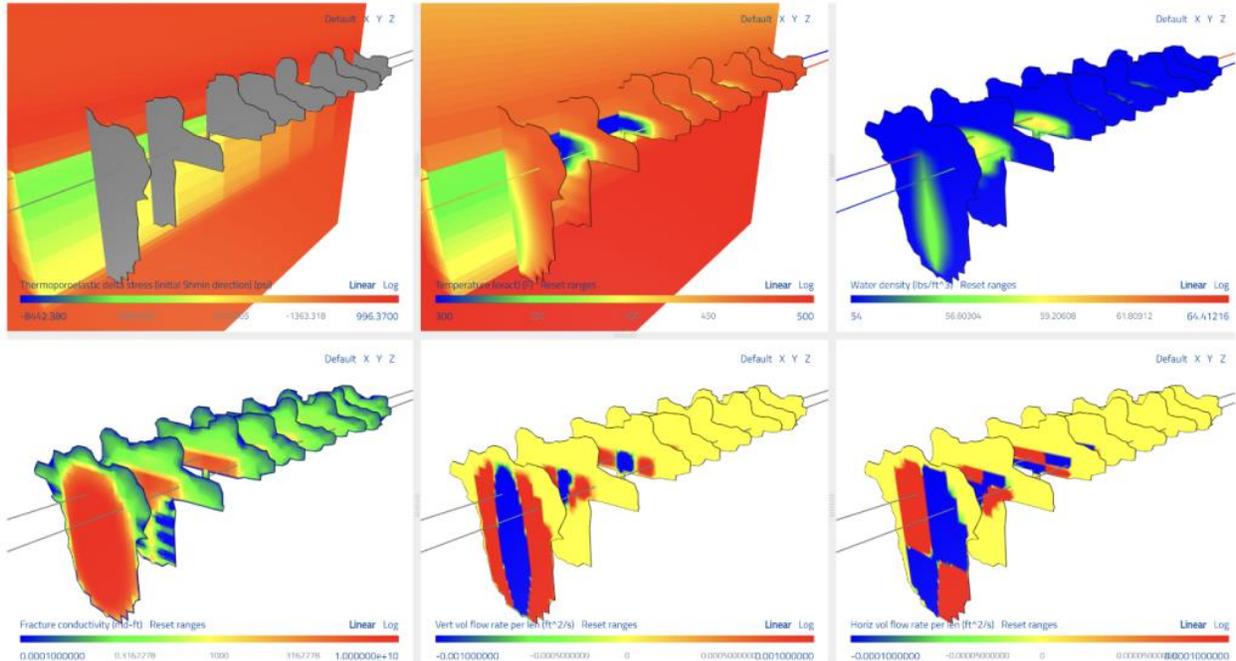
Figure 3: Flow distribution during long-term circulation with porothermoelastic stresses 'turned on'.

If the cracks were not open, we would expect density-driven buoyancy effects to be slight – the hydrostatic head difference caused by difference in water density are relatively small and should be much smaller than the pressure needed to push fluid through closed, finite conductivity fractures. But in the regions of the crack that are mechanically open due to cooling, the situation changes, and buoyancy becomes the dominant effect. Over 1000 ft of vertical distance, a 15% difference in water density corresponds to 65 psi. That is not much – but it is sufficient to drive flow in a wide-open crack.

Remarkably, from year 18 onward, the produced temperature *goes back up*. This happens as open fractures propagate downward from the injection well and access hotter and hotter rock. In addition, the 'opening front' of more and more fractures reaches the production well, improving the uniformity of flow. At year 30, the electricity generation rate is nearly 7 MWe, much better than in the simulation without porothermoelastic stresses.

This is a pretty extraordinary result, and potentially really good news for EGS. The downward propagation of fractures over years grows the stimulated rock volume over time and accesses hotter rock.

Do I believe this modeling result? The physical principles are sound – this is a process that will happen during long-term EGS circulation. But is there anything that makes me nervous that this model may be too optimistic?

The biggest risk occurs at the point when the fracture opening front first reaches the production well (during year 3 in the simulation above). If one or few fractures along the lateral became too dominant because of breakthrough of the 'crack opening front,' this could cause premature thermal breakthrough. A premature breakthrough could be mitigated by a cement squeeze or other intervention, but certainly, we would like to designs systems to minimize this risk.

Calculations suggest that with sufficient cooling, fractures will begin to propagate in the direction perpendicular to the orientation of the primary hydraulic fractures (Tarasovs and Ghassemi, 2010). This would be interesting to consider in future modeling, and the overall consequences on sweep efficiency are uncertain.

# Long-term circulation in a simulation with porothermoelastic stress changes and inflow control

Is there anything that we could do to mitigate the risk of this runaway thermal breakthrough scenario? The obvious – but costly – strategy would be to reduce circulation rate. Of course, this will reduce power generation and revenue. Another strategy would be to increase well-spacing. But if we space too far, this could reduce the number of 'connected fractures' between the wells.

A diversity of clever ideas are being pursued in the geothermal community for mitigating thermal breakthrough. For example, the US DOE is funding several projects on this topic.

In this section, I investigate the effect of physical 'inflow control' at the production well. Inflow control could be as simple as a cased well with very small 'perforation holes.' This approach would naturally prevent excessive flow location and would be a major safeguard against runaway thermal breakthrough.

Hypothetically, let's suppose that we case and cement the production well. Instead of fracturing it, we place a single tiny perforation shot – diameter of 0.1 inches – every 25 ft along the well. The pressure drop required to flow through these orifices would prevent flow localization. The pressure drop for flow through a circular orifice of diameter $D$ is (in metric units) (Cramer et al., 2019):

$$\Delta P = \frac{0.808 Q^2 \rho}{C^2 D^4}$$

Where $\rho$ is fluid density, $Q$ is volumetric flow rate, $D$ is diameter, and $C$ is coefficient of discharge (usually between 0.65 and 0.9).

For a diameter of 0.1 inches and a flow rate of 0.3 bbl/min, the value of ΔP will be large – 2700 psi. This pressure drop scales with the square of the flow rate, and so at 0.1 bbl/min, it will be only 300 psi. Effectively, it will be relatively easy to flow at 0.1 bbl/min; hard to flow at 0.3 bbl/min; and nearly impossible to flow at any rate higher than 0.3 bbl/min.

Would this design be technically feasible? Yes, with a bit of creativity. I am unsure whether perforation guns could be used to create such small diameter holes. Even if this was possible, it would be difficult to achieve uniform hole sizes. Probably, instead, we'd want to machine holes into the casing before placing it downhole. The holes would need to be closed during cementing, and so the design would require sliding sleeves that are opened after cementing.

After cementing, we'd need to 'break down' these flow points to create flow connections through the cemented annular region. With 320 holes along an 8000 ft lateral, injection at 96 bbl/min would imply 0.3 bbl/min per hole, achieving 2700 psi of 'perforation pressure drop' per shot and ensuring a uniform flow distribution. In other words, it would likely be possible to pump from the wellhead at 90-100 bbl/min and achieve breakdown at nearly every hole. The goal of breakdown would be to generate finite conductivity longitudinal cracks running along the annular region of the wellbore, outside the casing (for example, refer to Figure 5 from Ugueto et al., 2019). This would enable the cased/cemented production well to connect with the fractures created by the stimulation of the offset well, even if the fractures do not intersect the production well precisely at the location of the entry points.

There are a variety of other practical considerations, beyond the scope of this article. For example, a potential risk for small hole sizes is that they could be more vulnerable to plugging/screening from mobilized proppant during production. I expect that the sliding sleeve companies could improve on the general idea that I am outlining here. Indeed, there are already sophisticated solutions available for inflow control devices that are used in settings such as SAGD.

Why can't the limited entry used during fracturing help with flow control during circulation? During a typical plug-and-perf fracturing treatment in the injection well, injection may be performed at 80 bpm into eight perforation clusters (10 bpm per cluster) with diameter of 0.4'' and two shots per cluster. That implies a pressure drop at each cluster of 2300 psi – a strong limited-entry design that will enforce good uniformity of flow.

During long-term circulation, things are very different. Instead of injecting into a single stage, we are injecting into the entire lateral simultaneously. With 320 perf clusters, then even at the high flow rate of 60,000 bbl/day (42 bpm), the average flow rate per cluster is only 0.13 bpm. With the limited-entry design described above, that works out to 0.4 psi – negligible.

Thus, there is not a limited entry design that is suitable for both multistage hydraulic fracturing *and* flow control during long-term circulation. But it would be possible to use different designs in the production and injection wells – one optimized for fracturing and the other optimized for circulation. Or, a bit more complex, it could be possible to use sliding sleeves to use one set of holes for fracturing, and then shift them to use a different set of holes for long-term circulation.

Figure 4 shows the results from a simulation using the inflow control design in the production well (a single 0.1'' hole every 25 ft). The top left panel has a plot of production rate, temperature, and electrical generation over time. For the first ten years, production rate increases as thermoelastic cooling opens up the fractures, and produced temperature drops from 475˚ F to 400˚ F. Electricity generation starts at 5 MWe, and rises to 7-9 MWe within ten years. From year 10 onward, produced temperature and electricity generation *increase* over time. The downward propagation of the fractures accesses additional, hotter rock, causing a gradual recovery in produced temperature. The electricity generation exceeds 10 MWe during year 30 of operation, with no signs of abating!

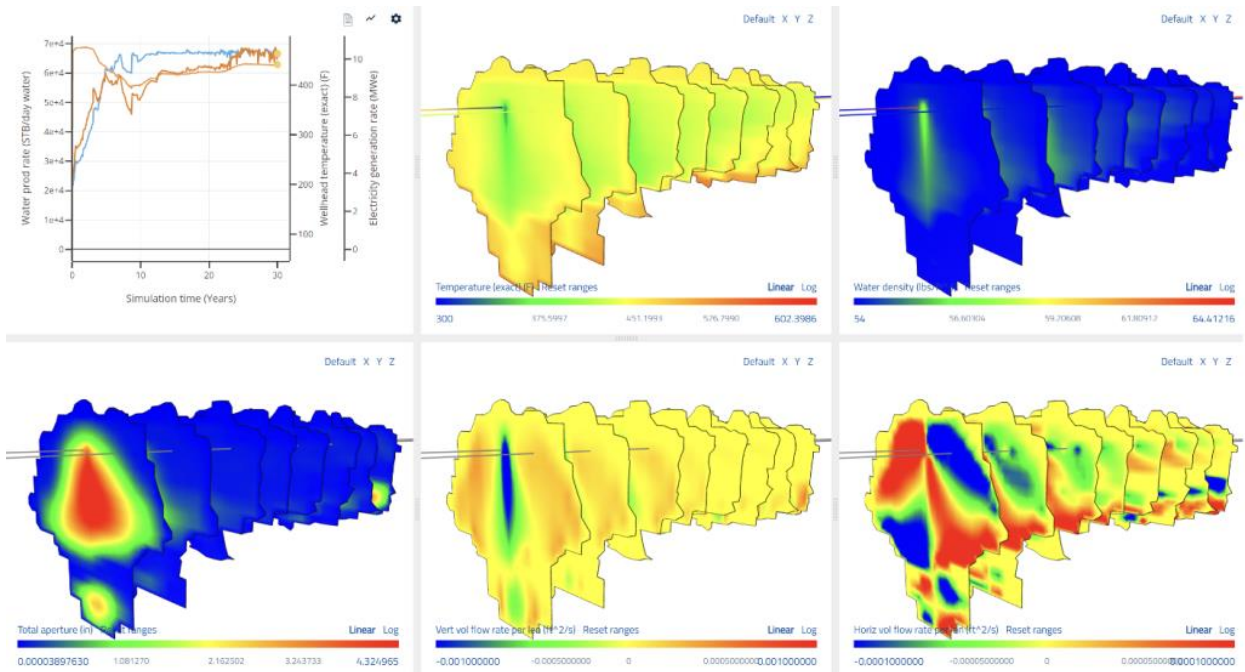

Figure 4: Long-term circulation with porothermoelastic stresses 'turned on' and flow-control in the production well to achieve a uniform flow distribution.

This result seems almost too good to be true; but really – this is a reasonable simulation result, reached without using any unreasonable model assumptions. The only 'reach' model assumption is that it depends on having good inflow-control capability in the production well.

Buoyancy-drive convection is the key process that extends the thermal longevity of the system. In Figure 5, I reran the simulation with a special setting to make water density unaffected by temperature. In this case, there is not any buoyancy-driven convection, and the fluid flows directly between the injection and production well, without sweeping through the rest of the crack. In this scenario, the produced temperature falls off significantly and never recovers – it is approaching 300˚ F by the end of the simulation, with only 2 MWe of electricity.

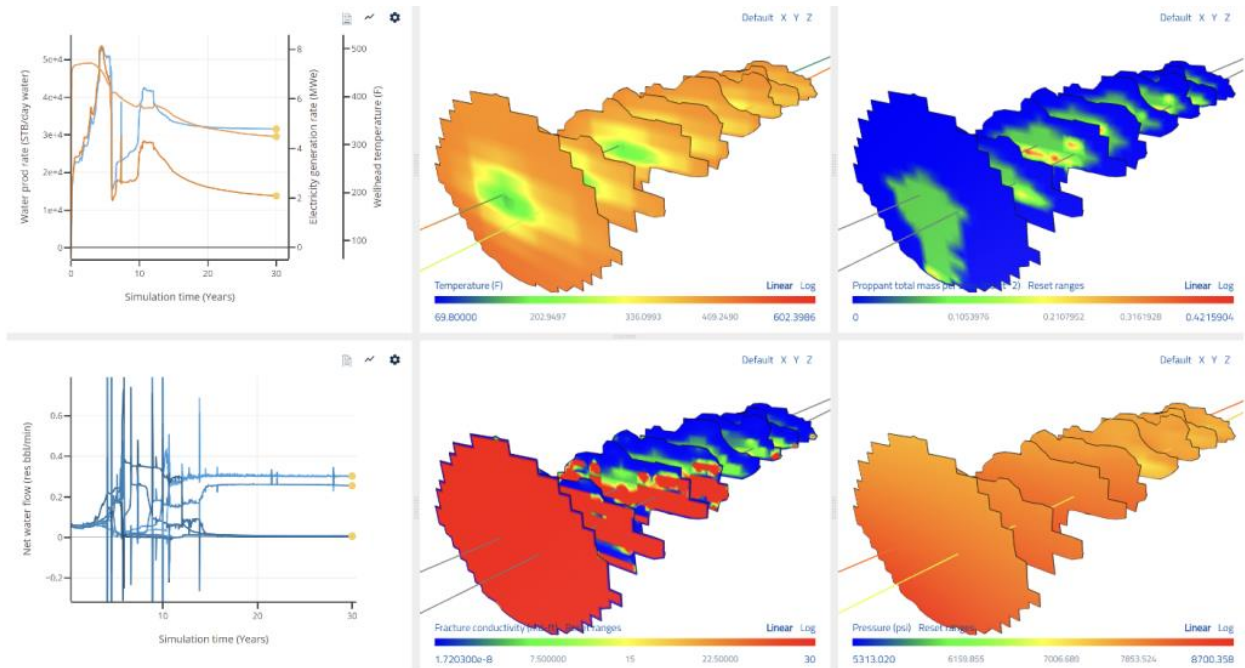

Figure 5: Long-term circulation with porothermoelastic stresses 'turned on,' flow-control in the production well to achieve a uniform flow distribution, and water density unaffected by temperature.

Is there anything we could do to *enhance* this tendency for buoyancy-driven convection? Hypothetically, if CO2 was used as the working fluid, instead of water, then the thermally-induced density difference would be greater, and we would see an even stronger effect.

# Conclusions

1. As predicted by modeling and demonstrated by recent field trials, multistage fracturing can enable high flow rates across a distributed network of flowing fractures (Norbeck and Latimer, 2023).
2. Thermal longevity is critical for economic performance. Classically, thermal breakthrough has been predicted based on calculations neglecting thermoelastic stress changes and other nonlinearities (Gringarten et al., 1975; Doe and McLaren, 2016). However, simulations considering thermoelastic stress reduction, fracture mechanics, and buoyancy yield complex and surprising effects.
3. Thermoelastic stress reduction and crack opening tends to accelerate thermal decline by making flow more uneven. However, it also increases injectivity. Furthermore, temperature-driven density differences drive convective circulation in the mechanically open regions of the fractures, significantly improving heat sweep efficiency. Buoyant circulation couples with thermoelastic stress reduction to drive downward crack propagation, improving the thermal longevity of the system.
4. When the first 'fracture opening fronts' reach the production well, there is a risk for rapid reduction of produced temperature. Inflow control configurations promise to prevent this from occurring, significantly improving thermal longevity and reducing project risk.

5. Because of buoyancy-driven convection and thermoelastic stress reduction and crack propagation, the simulation with inflow control in the production well exhibits outstanding reservoir performance – high flow rate and high produced temperature for more than 30 years. Designs without inflow control show strong – albeit somewhat lower – performance, but carry more risk of severe thermal breakthrough.